\documentclass[12pt]{article}
\usepackage{amsmath}
\usepackage{graphicx}
\usepackage{color}
\begin{document}
\baselineskip=18 pt
\begin{center}
{\large{\bf Relativistic quantum dynamics of spin-$0$ massive charged particle in the presence of external fields in 4D curved space-time with a cosmic string}}
\end{center}

\vspace{.5cm}

\begin{center}
{\bf Faizuddin Ahmed}\footnote{faizuddinahmed15@gmail.com ; faiz4U.enter@rediffmail.com}\\ 
{\it Ajmal College of Arts and Science, Dhubri-783324, Assam, India}
\end{center}

\vspace{.5cm}

\begin{abstract}

In this work, we investigate the relativistic quantum dynamics of spin-$0$ massive charged particles in a 4D curved space-time, the generalization of a cosmic string space-time. We investigate the Klein-Gordon equation in the presence of external fields in the considered framework. We also investigate a quantum particle described by the Klein-Gordon oscillator in the curved background space-time in the presence of external fields. We see that the energy eigenvalues get modifies in comparison to the results obtained in a cosmic string space-time.

\end{abstract}

{\it keywords:} Exact solutions, Relativistic wave equations, electromagnetic fields, topological defects, Solutions of Wave equations: bound states, special functions : Heun's differential equation \& Nikiforv-Uvarov method, energy spectrum, eigenfunctions.

\vspace{0.1cm}


\section{Introduction}

In recent years, study of relativistic wave equations, particularly the Klein-Gordon equation has attracted the attention of many authors because the solutions of this equation play an important role in obtaining the relativistic effects. The famous Klein-Gordon equation is one of the most frequently used equation that describes spin-$0$ particles such as mesons, bosons. The relativistic wave equations have been of growing interest for theoretical physicists in different branches of physics \cite{AWT,BT} including in nuclear and high energy physics (HEP) \cite{TYW,WG}. The exact solutions of the relativistic wave equations (non-relativistic or relativistic limit) contain all the necessary information needed for the complete description of quantum state of the system under consideration. However, analytical solutions are possible only in few simple cases such as the hydrogen atom, and the harmonic oscillator \cite{Landau}. In order to obtain the exact and approximate solutions of the relativistic wave equations, various quantum mechanical techniques have been employed ({\it e. g.}, \cite{AF,FC,CS4,HC}).

According to modern concepts of theoretical physics, different types of topological defects may be produced by the vacuum phase transition in the early Universe \cite{AV2,TWBK}. These include domain walls, cosmic strings and monopoles. The cosmic strings are linear defect, and the space-time produced by an idealized cosmic string is locally flat, however, globally conical, with a planar angle deficit determined by the string tension. Due to this conical singularity structure, a charged particle placed at rest in the cosmic string space-time becomes subjected to a repulsive electrostatic self-interaction \cite{BL,AGS} proportional to the inverse of the polar distance from the string. Also it was shown in \cite{ERBM} that the linear electric and magnetic sources in the space-time of cosmic string become subject to induced self-interactions \cite{MA2}. Cosmic strings is predicted in extensions of the standard model of particle physics \cite{AV3,MH}. Cosmic strings are also predicted in the framework of string theory \cite{MH2}.

The cosmic string space-time in polar coordinates $(t, r, \phi, z)$ is described by the following line element \cite{AV2,AV3,MH,AV,WAH,BL2,ERFM,LBC2} :
\begin{equation}
ds^2= - dt^2 + dr^2 +\eta^2\,r^2\,d\phi^2 + dz^2,
\label{1}
\end{equation}
where $\eta=1-4\mu $ is the topological parameter with $\mu$ being the linear mass density of the cosmic string.  In the cosmic string space-time, the parameter $\mu$ assumes values in the interval $ 0 < \mu < 1$ within the general relativity \cite{MOK,CF}. Furthermore, in the cylindrical symmetry we have that $ 0 < r < \infty$, $ 0 \leq \phi \leq  2\pi$ and $-\infty < z < \infty$. Several authors have been dealt with curved space-times in the presence of cosmic strings by investigating the influence of topological defects on the physical systems ({\it e. g.} \cite{MA,MGG,DVG,ERBDM,SGF}).

The analysis of a quantum mechanical system with gravitational interactions have recently been attracted a great deal of attention and an active field of research ({\it e. g.}, \cite{KBCF,MH3,KBak,FMA,LBC,GAM}). The analysis of a space-time produced by topological defects on the energy spectrum of the hydrogen atom, have been considered under both non-relativistic \cite{BL} and relativistic \cite{GAM2} limit. The Klein-Gordon equation with vector and scalar potentials of Coulomb-type under the influence of non-inertial effects in a cosmic string space-time, were investigated \cite{Santos}. In \cite{CFurtado}, motion of electrons in a uniform external magnetic field in the presence of screw dislocation, were investigated. In \cite{ALSN}, dynamics of quantum particle in the presence of a density of screw dislocations, were analyzed. In \cite{KBCF2}, non-inertial effects of rotating frames on the Landau quantization for neutral particles with a permanent magnetic dipole moment in the presence of topological defect, were studied. In \cite{HH}, the Klein-Gordon oscillator subject to a Cornell potential in a cosmic string space-time, were investigated. In \cite{HH2}, the Klein-Gordon fields in a spinning cosmic string space-time with a Cornell potential, were investigated. Other works in the cosmic string space-time would be in \cite{HH3,HH7,HH8,MS,MdM,HH9}. In addition, the relativistic wave equations are also investigated in G\"{o}del-type space-times ({\it e. g.}, see \cite{EPJC,EPJC2,EPJC3,EPJC4} and references therein).

This paper comprises as follow: analysis of a four-dimensional Petrov type D curved space-time with the stress-energy tensor in {\it section 2}; quantum dynamics of spin-$0$ massive charged particle in {\it sub-section 2.1}; Klein-Gordon oscillator in the presence of external fields in {\it sub-section 2.2}; and finally the conclusions in {\it section 3}.

\section{Analysis of (1+3)-dimensions type D curved space-time}

Consider the following static and cylindrical symmetry metric in polar coordinates $(t, r, \phi, z)$ given by 
\begin{equation}
ds^2=-dt^2+e^{-\frac{\alpha^2\,r^2}{2}}\,(dr^2+\eta^2\,r^2\,d\phi^2)+dz^2,
\label{2}
\end{equation}
where $\alpha >0 $ is an arbitrary parameter. For $\alpha\rightarrow 0$, the line-element (\ref{2}) reduces to cosmic string space-time (\ref{1}). Again for $\alpha \rightarrow 0$ and $\eta \rightarrow 1$, the study space-time reduces to Minkowski flat space metric in cylindrical coordinates. The non-zero component of the Riemann tensor $R_{\mu\nu\rho\sigma}$ for the metric (\ref{2}) is
\begin{equation}
R_{r\phi r\phi}=\eta^2\,\alpha^2\,r^2\,e^{-\frac{\alpha^2\,r^2}{2}}.
\label{3}
\end{equation}
The scalar curvature of the study space-time is 
\begin{equation}
R=R^{\mu}_{\mu}=2\,\alpha^2\,e^{\frac{\alpha^2\,r^2}{2}}.
\label{4}
\end{equation}
The metric tensor for the space-time (\ref{2}) to be 
\begin{equation}
g_{\mu\nu} ({\bf x})=\left (\begin{array}{llll}
-1 & \quad 0 & \quad 0 & 0 \\
\quad 0 & e^{-\frac{\alpha^2\,r^2}{2}} & \quad 0 & 0 \\
\quad 0 & \quad 0 & \frac{\eta^2\,r^2}{e^{\frac{\alpha^2\,r^2}{2}}} & 0 \\
\quad 0 & \quad 0 & \quad 0 & 1
\end{array} \right)
\label{5}
\end{equation}
with its inverse 
\begin{equation}
g^{\mu\nu} ({\bf x})=\left (\begin{array}{llll}
-1 & 0 & 0 & 0 \\
\quad 0 & e^{\frac{\alpha^2\,r^2}{2}} & 0 & 0 \\
\quad 0 & 0 & \frac{e^{\frac{\alpha^2\,r^2}{2}}}{\eta^{2}\,r^{2}} & 0 \\
\quad 0 & 0 & 0 & 1
\end{array} \right)
\label{6}
\end{equation}
The metric has signature $(-,+,+,+)$ and the determinant of the corresponding metric tensor $g_{\mu\nu}$ is
\begin{equation}
det\;g=-\eta^2\,r^2\,e^{-\alpha^2\,r^2}.
\label{7}
\end{equation}

The null tetards of the above space-time are
\begin{eqnarray}
&&k_{\mu}=\frac{1}{\sqrt{2}}\,(1,0,0,1),\nonumber\\
&&l_{\mu}=\frac{1}{\sqrt{2}}\,(1,0,0,-1),\nonumber\\
&&m_{\mu}=\frac{e^{-\frac{\alpha^2\,r^2}{4}}}{\sqrt{2}}\,(0,1,i\,\eta\,r,0),\nonumber\\
&&\bar{m}_{\mu}=\frac{e^{-\frac{\alpha^2\,r^2}{4}}}{\sqrt{2}}\,(0,1,-i\,\eta\,r,0).
\label{8}
\end{eqnarray}
The Einstein Field Equations are given by ($c=1=\hbar$, $8\,\pi\,G=1$)
\begin{equation}
G_{\mu\nu}=R_{\mu\nu}-\frac{1}{2}\,g_{\mu\nu}\,R=T_{\mu\nu},
\label{9}
\end{equation}
where $T_{\mu\nu}$ is the stress-energy tensor and we choose the following one
\begin{equation}
T_{\mu\nu}=\epsilon\,k_{\mu}\,k_{\nu}+(\rho+p)\,(k_{\mu}\,l_{\nu}+l_{\mu}\,k_{\nu})+p\,g_{\mu\nu},
\label{10}
\end{equation}
where $\epsilon$ as the radiation energy density, $\rho$ as the string energy density and $p$ as the string pressure \cite{SH,AZW,VH,LKP,SGG,RC} and the vector $k_{\mu}, l_{\mu}$ are null and orthogonal except $k^{\mu}\,l_{\mu}=-1$.

For $\epsilon=0$, the stress-energy tensor (\ref{10}) corresponds to Type I fluid \cite{SH}. For $\epsilon=0$ and $p=0$, it represents string dust \cite{LKP,SGG}. Therefore, for string dust we have
\begin{equation}
T_{\mu\nu}=\rho\,(k_{\mu}\,l_{\nu}+l_{\mu}\,k_{\nu}),
\label{11}
\end{equation}
where 
\begin{equation}
T_{\mu\nu}\,k^{\mu}\,l^{\nu}=\rho=T_{\mu\nu}\,k^{\nu}\,l^{\mu}\quad,\quad T^{\mu}_{\mu}=T=-2\,\rho.
\label{12}
\end{equation}

The non-zero components of the Einstein tensor $G_{\mu\nu}$ for the metric (\ref{2}) are
\begin{equation}
G_{tt}=-G_{zz}=\alpha^2\,e^{\frac{\alpha^2\,r^2}{2}}.
\label{13}
\end{equation}
From the field equations (\ref{9}) using (\ref{8}), (\ref{10}) and (\ref{13}), we have
\begin{equation}
\epsilon=0\quad,\quad p=0\quad,\quad \rho=\alpha^2\,e^{\frac{\alpha^2\,r^2}{2}}.
\label{14}
\end{equation}
Thus the stress energy tensor is of Type I fluid with zero pressure, $p=0$, called string dust which obeys the weak as well as the other energy conditions \cite{SH}.

The non-zero component of the Weyl scalars for the space-time (\ref{2}) is 
\begin{equation}
\Psi_2=-\frac{\alpha^2}{6}\,e^{\frac{\alpha^2\,r^2}{2}},
\label{15}
\end{equation}
while the others are $\Psi_0=\Psi_1=0=\Psi_3=\Psi_4$. This clearly shows that the study space-time is of type D metric in the Petrov classification scheme. The Kretschamnn scalar of the study space-time is
\begin{equation}
K=R^{\mu\nu\rho\sigma}\,R_{\mu\nu\rho\sigma}=4\,\alpha^4\,e^{\alpha^2\,r^2}
\label{16}
\end{equation}
We can see that the string dust energy density $\rho$, the Weyl scalar $\Psi_2$ and the Kretschmann scalar $K$ are finite at $r \rightarrow 0$. We can see that the stress-energy tensor of the study space-time is of zero pressure Type I fluid called string dust which satisfy all the energy conditions.

\subsection{Spin-$0$ massive charged particle in the presence of external fields in a curved space-time}

The relativistic quantum dynamics of  scalar particles of mass $m$ is described by the Klein-Gordon (KG) equation
\begin{equation}
\frac{1}{\sqrt{-g}}\,\partial_{\mu} (\sqrt{-g}\,g^{\mu\nu}\,\partial_{\nu}\,\Psi)=m^2\,\Psi,
\label{aa1}
\end{equation}
with $g$ is the determinant of metric tensor with $g^{\mu\nu}$ its inverse.

Using the space-time (\ref{2}) into (\ref{aa1}), we get the following equation:
\begin{equation}
(-\partial_{t}^2+\partial_{z}^2)\,\Psi+e^{\frac{\alpha^2\,r^2}{2}}[\frac{1}{r}\,\Psi' (r)+\Psi'' (r)+\frac{1}{\eta^2 r^2}\,\partial_{\phi}^2\,\Psi]=m^2\,\Psi.
\label{aa2}
\end{equation}

We now introduce an electromagnetic interactions into the Klein-Gordon equation through the minimal substitution \cite{ERFM,Bir,LCNS}
\begin{equation}
\partial_{\mu}\rightarrow D_{\mu}=\partial_{\mu}-i\,e\,A_{\mu},
\label{aa3}
\end{equation}
where $e$ is the electric charge, and $A_{\mu}=(0,\vec{A})$ is the electromagnetic four-vector potential.

The electromagnetic vector potential associated with a uniform external magnetic field in Coulomb-gauge is given by
\begin{equation}
\vec{A}=(0,A_{\phi},0),\quad A_{\phi}=-\frac{1}{2}\,\eta\,B_{0}\,r^2
\label{aa4}
\end{equation}
such that the magnetic field parallel to the $z$-axis is 
\begin{equation}
\vec{B}=\vec{\nabla} \times \vec{A}=-B_{0}\,\hat{k}.
\label{aa5}
\end{equation}
Therefore, the KG-equation (\ref{aa2}) becomes
\begin{equation}
(-\partial_{t}^2+\partial_{z}^2-m^2)\,\Psi+e^{\frac{\alpha^2\,r^2}{2}}[\Psi'' (r)+\frac{1}{r}\,\Psi' (r)+\frac{1}{\eta^2\,r^2}\,(\partial_{\phi}-i\,e\,A_{\phi})^2\,\Psi]=0.
\label{aa6}
\end{equation}

The equation (\ref{aa6}) is independent of $t,\phi, z$, so it is appropriate to choose the following ansatz for the function $\Psi$
\begin{equation}
\Psi (t,r,\phi,z)=e^{i\,(-E\,t+l\,\phi+k\,z)}\,\psi (r),
\label{aa7}
\end{equation}
where $E=i\,\partial_{t}$ is the total energy, $l=0,\pm\,1,\pm\,2....\in {\bf Z}$ is the eigenvalues of the $z$-component of the angular momentum operator, and $ -\infty < k < \infty $ is the eigenvalues of the $z$-component of the linear momentum operator.

Substituting the ansatz (\ref{aa7}) into the Eq. (\ref{aa6}), we obtain the following differential equation for $\psi (r)$:
\begin{equation}
\psi''(r)+\frac{1}{r}\,\psi' (r)+[f(r)-\frac{{\bar l}^2}{r^2}-m^2\,\omega^2_{c}\,r^2-2\,m\,\omega_{c}\,{\bar l}]\,\psi (r)=0,
\label{aa8}
\end{equation}
where the function $f(r)$ is
\begin{equation}
f(r)=e^{-\frac{\alpha^2\,r^2}{2}}(E^2-k^2-m^2).
\label{aa9}
\end{equation}
And
\begin{equation}
{\bar l}=\frac{l}{\eta}\quad,\quad \omega_c=\frac{e\,B_{0}}{2\,m}
\label{aa10}
\end{equation}
respectively are the effective angular momentum and cyclotron frequency of the particle.

Now, we consider an approximation on the exponential term in the function $f(r)$. Setting $Z=\frac{\alpha^2\,r^2}{2}$ and apply the binomial theorem up to the power of $Z^3$ (that means $r^{6}$ for mathematical simplicity), we have
\begin{equation}
e^{-Z}=1-Z+\frac{Z^2}{2}-\frac{Z^3}{6}
\label{aa11}
\end{equation}
which implies
\begin{equation}
e^{-\frac{\alpha^2\,r^2}{2}}=1-\frac{\alpha^2\,r^2}{2}+\frac{\alpha^4\,r^4}{8}-\frac{\alpha^6\,r^6}{48}.
\label{aa12}
\end{equation}
Therefore the function $f (r)$ can be express as
\begin{equation}
f(r)= e^{-\frac{\alpha^2\,r^2}{2}}(E^2-k^2-m^2)=a+b\,r^2+c\,r^4+d\,r^6,
\label{aa13}
\end{equation}
where
\begin{eqnarray}
&&a=(E^2-k^2-m^2),\nonumber\\
&&b=(E^2-k^2-m^2)(-\frac{\alpha^2}{2})=-\frac{a\,\alpha^2}{2},\nonumber\\
&&c=(E^2-k^2-m^2)(\frac{\alpha^4}{8})=\frac{a\,\alpha^4}{8},\nonumber\\
&&d=(E^2-k^2-m^2)(-\frac{\alpha^6}{48})=-\frac{a\,\alpha^6}{48}.
\label{aa14}
\end{eqnarray}

\vspace{0.1cm}
{\bf Case 1}: We choose the function $f(r) = a+b\,r^2+c\,r^4+d\,r^6$.
\vspace{0.1cm}

Using the above function $f(r)$ into the radial equation Eq. (\ref{aa8}), we have
\begin{equation}
\psi'' (r)+\frac{1}{r}\,\psi' (r)+[a_0+b_0\,r^2+c\,r^4+d\,r^6-\frac{{\bar l}^2}{r^2}]\,\psi (r)=0,
\label{aa15}
\end{equation}
where
\begin{eqnarray}
&&a_0=a-2\,m\,\omega_c\,{\bar l},\nonumber\\
&&b_{0}=b-m^2\,\omega^2_{c}.
\label{aa16}
\end{eqnarray}

Substituting the following
\begin{equation}
\psi (r)=\frac{U (r)}{\sqrt{r}}
\label{aa17}
\end{equation}
into the Eq. (\ref{aa15}), we get
\begin{equation}
U'' (r)+[a_0+b_0\,r^2+c\,r^4+d\,r^6+\frac{(-{\bar l}^2+\frac{1}{4})}{r^2}]\,U (r)=0.
\label{aa18}
\end{equation}
Changing the variable $r^2=x$ in the Eq. (\ref{aa18}), we have
\begin{equation}
U'' (x)+\frac{U' (x)}{2\,x}+[\frac{\frac{a_0}{4}}{x}+\frac{b_0}{4}+\frac{c}{4}\,x+\frac{d}{4}\,x^2+\frac{\frac{1}{4}(-{\bar l}^2+\frac{1}{4})}{x^2}]\,U (x)=0.
\label{aa19}
\end{equation}
Again substituting the following 
\begin{equation}
U (x)=\frac{\Phi (x)}{x^{\frac{1}{4}}}
\label{aa20}
\end{equation}
into the Eq. (\ref{aa19}), we get
\begin{equation}
\Phi'' (x)+[\frac{\frac{a_0}{4}}{x}+\frac{b_0}{4}+\frac{c}{4}\,x+\frac{d}{4}\,x^2+\frac{\frac{1}{4} (-{\bar l} ^2 + 1)}{x^2}]\,\Phi (x)=0.
\label{aa21}
\end{equation}

Now, we use the appropriate boundary conditions to investigate the bound states solutions in this problem. It is require that the wave functions must be regular both at $x \rightarrow 0$ and $x\rightarrow \infty$. These conditions are necessary since the wave functions must be well-behaved in these limit. Let us impose the requirement that $\Phi (x) \rightarrow 0$ when $x \rightarrow 0$ and $x\rightarrow \infty$. Suppose the possible solutions to the equation (\ref{aa21}) can be express in terms of an unknown function H(x) as follows:
\begin{equation}
\Phi (x)=x^{A}\,e^{-(B\,x + D\,x^2)}\,H(x).
\label{aa22}
\end{equation}
Here $A, B, D$ have to be determined now. Substituting Eq. (\ref{aa22}) into the Eq. (\ref{aa21}), we get
\begin{eqnarray}
&&H''(x)+ [\frac{2A}{x}-2\,B-4\,D\,x] H' (x) +[\frac{A^2-A+\frac{1}{4}(-{\bar l}^2+1)}{x^2} +\frac{\frac{a_0}{4}- 2\,A\,B}{x}\nonumber\\
&&+\frac{b_0}{4}+B^2-2\,D\,(1+2\,A) +(\frac{c}{4}+4\,B\,D)\,x+(\frac{d}{4}+4\,D^2)\,x^2]\,H(x)=0.
\label{aa23}
\end{eqnarray}
Now we set the coefficients of $x, x^2, x^{-2}$ equals to zero in the above equation, we get (taking $+$ sign):
\begin{eqnarray}
&&A=\frac{1}{2} (1+{\bar l}),\nonumber\\ 
&&D=\frac{\sqrt{-d}}{4},\nonumber\\
&&B=-\frac{c}{4\sqrt{-d}}.
\label{aa24}
\end{eqnarray}
With these, the Eq. (\ref{aa23}) can now be written as
\begin{eqnarray}
&&H''(x) + [\frac{2\,A}{x} -2\,B-4\,D\,x]\,H' (x)+[\frac{\frac{a_0}{4}-2\,A\,B}{x}+\frac{b_0}{4}+B^2\nonumber\\
&&-2\,D \,(1+2 A)]\,H(x)=0.
\label{aa25}
\end{eqnarray}
The above Eq. (\ref{aa25}) can be express as
\begin{equation}
\frac{d^2\,H}{dx^2}+[\frac{\gamma}{x}-\delta-\epsilon\,x]\,\frac{d H}{dx}+ [\beta -\frac {q}{x}] \,H(x)=0,
\label{aa26}
\end{equation}
where
\begin{eqnarray}
&&\gamma=2\,A,\nonumber\\
&&\delta=2\,B,\nonumber\\
&&\epsilon=4\,D,\nonumber\\
&&q=(-\frac{a_0}{4}+2\,A\,B),\nonumber\\
&&\beta=\frac{b_0}{4}+B^2-2\,D\,(1+2\,A).
\label{aa27}
\end{eqnarray}
Eq. (\ref{aa26}) is the biconfluent Heun's differential equation \cite{ERFM,SYS,AR} and $H(x)$ is the Heun polynomials. 

Let us use the Frobenius method to find the solution. The solution to Eq. (\ref{aa26}) can be written as a power series expansion around the origin \cite{GBA}:
\begin{equation}
H (x) = \sum^{\infty}_{i=0} c_{i}\,x^{i}.
\label{aa28}
\end{equation}
Substituting Eq. (\ref{aa28}) into the Eq. (\ref{aa26}), we get the following recurrence relation for the coefficients:
\begin{equation}
c_{n+2}=\frac{1}{(n+2)(n+1+\gamma)}\,[\{q+\delta (n+1) \}\,c_{n+1}-(\beta-\epsilon\,n)\,c_{n}].
\label{aa29}
\end{equation}
And the various coefficients are
\begin{equation}
c_1=\frac{q}{\gamma}\,c_0,\quad c_2=\frac{1}{2(1+\gamma)}\,[(q+\delta)\,c_{1}-\beta\,c_{0}].
\label{aa30}
\end{equation}

Since the wave-functions $\Phi $ has no divergence at $ x \rightarrow 0$ and $x \rightarrow \infty$ and we have written the function $H (x)$ as a power series expansion around the origin (\ref{aa28}). Through the recurrence relation (\ref{aa29}), one can see that the power series expansion $H (x)$ becomes a polynomial of degree $n$ by imposing the following two conditions \cite{ERFM,AVV,JM}:
\begin{eqnarray}
\label{aa31}
&&\beta-\epsilon\,n=0,\quad (n=1,2,3,4,...),\\
\label{aa32}
&&c_{n+1}=0.
\end{eqnarray}
Using the condition $\beta=\epsilon\,n$, we have 
\begin{equation}
\frac{b}{4}-\frac{1}{4}\,m^2\,\omega^2_{c}+B^2-2\,D\,(1+2 A) = 4\,n\,D.
\label{aa33}
\end{equation}
Substituting $A, B, D$ using Eq. (\ref{aa24}) into the condition (\ref{aa31}), we get the following second degree energy eigenvalues (taking $+$ sign)
\begin{equation}
E^{2}_{n,l}=k^2+m^2+\frac{16}{75}\,[\alpha\,(2\,n+2+\frac{l}{\eta})+\sqrt{\alpha^2\,(2\,n+2+\frac{l}{\eta})^2-\frac{15\,m^2\,\omega^2_{c}}{4\,\alpha^2}}]^2.
\label{aa34}
\end{equation}

The corresponding wave functions are given by
\begin{eqnarray}
\Phi_{n,l} (x)&=&x^{\frac{1}{2}(1+\frac{|l|}{\eta})}\,e^{\frac{x}{4\,\sqrt{-d}}\,(c+d\,x)}\,H (x),\nonumber\\
&=&x^{\frac{1}{2}(1+\frac{|l|}{\eta})}\,e^{ \frac{x\,\alpha}{8}\,(1-\frac{x\,\alpha^2}{6})\,\sqrt{3\,(E^{2}_{n,l}-k^2-m^2)}}\,H (x),
\label{aa35}
\end{eqnarray}
where $E_{n,l}$ is given in Eq. (\ref{aa34}). 

Now, we employing the additional recurrence condition $c_{n+1}=0$ to obtain the individual energy eigenvalues and corresponding wave functions one by one as done in Refs. \cite{AVV,JM}.

For $n=1$, we have $\beta=\epsilon$ and $c_2=0$ which implies
\begin{equation}
(q+\delta)\,c_{1}=\beta\,c_{0}\Rightarrow (q+\delta)\,\frac{q}{\gamma}\,c_0=\epsilon\,c_0\Rightarrow
\frac{q}{\gamma}=\frac{\epsilon}{q+\delta}
\label{17}
\end{equation}
a constraint on the physical parameters $(B_0, m, \alpha, \eta, l, k)$ and energy $E$. The energy eigenvalues for $n=1$ is 
\begin{equation}
E^{2}_{1,l}=k^2+m^2+\frac{16}{75}\,[\alpha\,(4+\frac{l}{\eta})+\sqrt{\alpha^2\,(4+\frac{l}{\eta})^2-\frac{15\,m^2\,\omega^2_{c}}{4\,\alpha^2}}]^2.
\label{18}
\end{equation}
The corresponding eigenfunctions
\begin{eqnarray}
\Phi_{1,l} (x)=x^{\frac{1}{2}(1+\frac{|l|}{\eta})}\,e^{ \frac{x\,\alpha}{8}\,(1-\frac{x\,\alpha^2}{6})\,\sqrt{3\,(E^{2}_{1,l}-k^2-m^2)}}\,[c_0+c_1\,x],
\label{19}
\end{eqnarray}
where 
\begin{equation}
c_1=[-2\,\sqrt{3\,(m^2+k^2-E^2_{1,l})}-\frac{(E^2_{1,l}-m^2-k^2-2\,m\,\omega_c\,{\bar l})}{4\,(1+{\bar l})}]\,c_0. 
\label{20}
\end{equation}
And $E_{1,l}$ is given in Eq. (\ref{18}).

Worthwhile it is better to mention that for $\alpha \rightarrow 0$, the study space-time (\ref{2}) reduces to the cosmic string space-time. In that case, from Eq. (\ref{aa34}) one will not get back the similar energy eigenvalues of a charged particle as obtained in Ref. \cite{ERFM}  directly. To obtain the eigenvalues for $\alpha \rightarrow 0$, we substitute $b=c=d=0$ into the Eq. (\ref{aa15}), we have
\begin{equation}
\psi'' (r)+\frac{1}{r}\,\psi' (r)+[a_0-m^2\,\omega^2_{c}\,r^2-\frac{{\bar l}^2}{r^2}]\,\psi (r)=0,
\label{aa36}
\end{equation} 
Transforming $m\,\omega_{c}\,r^2=x$ into the Eq. (\ref{aa36}), one will obtain the following equation \cite{AF}
\begin{equation}
\psi'' (x)+\frac{1}{x}\,\psi' (x)+\frac{1}{x^2} (-\xi_1\,x^2+\xi_2\,x-\xi_3)\,\psi (x)=0,
\label{aa37}
\end{equation} 
where
\begin{equation}
\xi_1=\frac{1}{4}\quad,\quad \xi_2=\frac{a}{4\,m\,\omega_{c}}\quad,\quad \xi_3=\frac{{\bar l}^2}{4}=\frac{l^2}{4\eta^2}.
\label{aa38}
\end{equation}
The energy eigenvalues is given by
\begin{equation}
E_{n,l}=\pm\sqrt{k^2+m^2+2\,m\,\omega_{c}\,(2n+1+\frac{|l|}{\eta}+\frac{l}{\eta})}
\label{aa40}
\end{equation}
which is similar to the eigenvalues obtained in Ref. \cite{ERFM} in a cosmic string space-time (see Eq. (23) in Ref. \cite{ERFM}).

Thus we can see that the energy eigenvalues (\ref{aa34}) of a charged particle in the presence of external fields in a curved space-time in comparison to the result in Ref. \cite{ERFM} in a cosmic string space-time get modifies.

\vspace{0.1cm}
{\bf Case 2} : We choose the function $f(r) =a+b\,r^2$ (assuming $\alpha < < 1$, thereby neglecting the higher order terms).
\vspace{0.1cm}

Substituting the above function into the Eq. (\ref{aa8}), we have 
\begin{equation}
\psi'' (r)+\frac{1}{r}\,\psi' (r)+(a_0+b_0\,r^2-\frac{{\bar l}^2}{r^2})\,\psi (r)=0.
\label{bb1}
\end{equation}
Transforming a new variable $r^2=x$ into the Eq. (\ref{bb1}), we have
\begin{equation}
\psi'' (x)+\frac{1}{x}\,\psi' (x)+\frac{1}{4\,x}(a_0+b_0\,x-\frac{{\bar l}^2}{x})\,\psi (x)=0.
\label{bb2}
\end{equation}
The above equation can be express as \cite{AF}
\begin{equation}
\psi'' (x)+\frac{1}{x}\,\psi' (x)+\frac{1}{x^2}\,(-\xi_{1}\,x^2+\xi_{2}\,x-\xi_3)\,\psi (x)=0,
\label{bb3}
\end{equation}
where
\begin{equation}
\xi_1=-\frac{b}{4}+\frac{1}{4}\,m^2\,\omega^2_{c}\quad,\quad \xi_2=\frac{a-2\,m\,\omega_{c}\,{\bar l}}{4}\quad,\quad \xi_3=\frac{{\bar l}^2}{4}.
\label{bb4}
\end{equation}
The energy eigenvalues is given by
\begin{eqnarray}
&&E^{2}_{n,l}=k^2+m^2+2\,m\,\omega_{c}\,\frac{l}{\eta}+\alpha^2\,(2\,n+1+\frac{|l|}{\eta})^2 \nonumber\\
&+&(2\,n+1+\frac{|l|}{\eta}) \sqrt{\alpha^4\,(2\,n+1+\frac{|l|}{\eta})^2+4\,m^2\,\omega^2_{c}+4\,m\,\omega_{c}\,\alpha^2\,\frac{l}{\eta}}.
\label{bb5}
\end{eqnarray}
For $\alpha \rightarrow 0$, the energy eigenvalues (\ref{bb5}) reduces to the result obtained in Ref. \cite{ERFM}.

Thus we can see that the energy eigenvalues (\ref{bb5}) in comparison to the result obtained in Ref. \cite{ERFM} in a cosmic string space-time get modifies (increases) due to the presence of the extra parameter $\alpha$ in the curved space-time.

 From the above analysis by {\bf Case 1}--{\bf Case 2}, one can see that the energy spectrum Eq. (\ref{aa34}) and (\ref{bb5}) of a charged particle in the presence of external fields in a curved space-time are the modified result in comparison to the case obtained in Ref. \cite{ERFM} in a cosmic string space-time. This modification of the energy eigenvalues is due to the extra parameter $\alpha$ which causes the cosmic string space-time (Riemannian flat space-time except conical singularity) into a curved one (non-flat Riemannian space-time).

\subsection{Klein-Gordon Oscillator in the presence of external fields in a curved space-time}

 Another system of interest that may be considered is the Klein-Gordon oscillator in the background of the curved space-time. The Klein-Gordon oscillator \cite{Bru,VVD} was inspired by earlier papers on the Dirac oscillator \cite{MM} applied to spin-$\frac{1}{2}$ particles. In recent years, several studies have addressed the Kelin-Gordon oscillator in relativistic quantum systems \cite{YJX,WJH,KB,RVM,RLLV,MLL,NAR,AB,ZW}. These sources of gravitational fields play an important role in condensed matter physics systems \cite{CS,AMMC,CS2,CS3}. Besides topological defects like cosmic strings \cite{AV} and domain walls \cite{AV2}, a global monopole \cite{MB2} provides a tiny relation between the effects in cosmology and gravitation and those in condensed matter physics systems, where topological defects analogous to cosmic strings appear in phase transitions in liquid crystals \cite{HM,FM}.

To couple the Klein-Gordon field with oscillator \cite{Bru,VVD,Mirza}, following change in the momentum operator is taken:
\begin{equation}
p_{\mu}\rightarrow p_{\mu}+i\,m\,\Omega\,X_{\mu},
\label{dd1}
\end{equation}
where $m$ is the particle mass at rest, $\Omega$ is the frequency of the oscillator and the vector $X_{\mu}$ is defined as
\begin{equation}
X_{\mu}=(0,r,0,0),
\label{dd2}
\end{equation}
with $r$ being the distance from the particle to the string. Thus, the Klein-Gordon equation for the oscillator becomes
\begin{equation}
\frac{1}{\sqrt{-g}}\,(\partial_{\mu}+m\,\Omega\,X_{\mu})\sqrt{-g}\,g^{\mu\nu}\,(\partial_{\nu}-m\,\Omega\,X_{\nu})\,\Psi=m^2\,\Psi.
\label{dd3}
\end{equation}

Using the space-time (\ref{2}) into the Eq. (\ref{dd3}), we get
\begin{eqnarray}
&&(-\partial^{2}_{t}+\partial^{2}_{z}-m^2)\,\Psi+e^{\frac{\alpha^2\,r^2}{2}}\,[\Psi'' (r)+\frac{1}{r}\,\Psi'(r)-2\,m\,\Omega\,\Psi-m^2\,\Omega^2\,r^2\,\Psi\nonumber\\
&&+\frac{1}{\eta^2\,r^2}\,\partial^2_{\phi}\,\Psi]=0.
\label{dd4}
\end{eqnarray}
We now introduce an external uniform magnetic field through the minimal coupling 
\begin{equation}
\partial_{\phi} \rightarrow \partial_{\phi}-i\,e\,A_{\phi}
\label{dd5}
\end{equation}
where $A_{\phi}$ is given in Eq. (\ref{aa4}). Therefore, Eq. (\ref{dd4}) becomes
\begin{eqnarray}
&&(-\partial^{2}_{t}+\partial^{2}_{z}-m^2)\,\Psi+e^{\frac{\alpha^2\,r^2}{2}}\,[\Psi'' (r)+\frac{1}{r}\,\Psi'(r)\nonumber\\
&&+\{-2\,m\,\Omega-m^2\,\Omega^2\,r^2+\frac{1}{\eta^2\,r^2}\,(\partial_{\phi}-i\,e\,A_{\phi})^2\}\,\Psi]=0.
\label{dd6}
\end{eqnarray}
Substituting the ansatz Eq. (\ref{aa7}) into the Eq. (\ref{dd6}), we get
\begin{equation}
\psi'' (r)+\frac{1}{r}\,\psi' (r)+[f(r)-2\,m\,\Omega-2\,m\,\omega_c\,{\bar l}-m^2\,(\Omega^2+\omega^2_{c})\,r^2-\frac{{\bar l}^2}{r^2}]\,\psi (r)=0,
\label{dd7}
\end{equation}
where the function $f (r)$ is defined earlier.

\vspace{0.1cm}
{\bf Case 1}: We choose the function $f(r) = a+b\,r^2+c\,r^4+d\,r^6$.
\vspace{0.1cm}

Using the above function $f(r)$ into the Eq. (\ref{dd7}), we get
\begin{equation}
\psi'' (r)+\frac{1}{r}\,\psi' (r)+[\tilde{a}+\tilde{b}\,r^2+c\,r^4+d\,r^6-\frac{{\bar l}^2}{r^2}]\,\psi (r)=0,
\label{dd8}
\end{equation}
where 
\begin{eqnarray}
&&\tilde{a}=a-2\,m\,\Omega-2\,m\,\omega_c\,{\bar l},\nonumber\\
&&\tilde{b}=b-m^2\,\Omega^2-m^2\,\omega^2_{c}.
\label{dd9}
\end{eqnarray}

Using the similar technique as done earlier, one can obtain the following biconfluent Heun's differential equation for the Eq. (\ref{dd8}) :
\begin{equation}
H'' (x) + [\frac{\gamma}{x}-\delta-\epsilon\,x]\,H' (x)+ [\tilde{\beta} -\frac {\tilde{q}}{x}]\,H(x)=0,
\label{dd10}
\end{equation}
where we defined the following 
\begin{eqnarray}
&&\gamma=2\,A,\nonumber\\
&&\delta=2\,B,\nonumber\\
&&\epsilon=4\,D,\nonumber\\
&&\tilde{\beta}=\frac{\tilde{b}}{4}+B^2-2\,D\,(1+2\,A),\nonumber\\
&&\tilde{q}=-\frac{\tilde{a}}{4}+2\,A\,B.
\label{dd11}
\end{eqnarray}
And $A, B, D$ are given earlier.

Substituting the series solution Eq. (\ref{aa28}) into the Eq. (\ref{dd11}), we get the following recurrernce relation for the coefficients:
\begin{equation}
c_{n+2}=\frac{1}{(n+2)(n+1+\gamma)}\,[\{ \tilde{q}+\delta\,(n+1) \}\,c_{n+1}-(\tilde{\beta}-\epsilon\,n)\,c_{n}].
\label{dd12}
\end{equation}
And the various coefficients are
\begin{eqnarray}
&&c_1=\frac{\tilde{q}}{\gamma}\,c_0,\nonumber\\
&&c_2=\frac{1}{2(1+\gamma)}\,[(\tilde{q}+\delta)\,c_{1}-\tilde{\beta}\,c_{0}].
\label{dd13}
\end{eqnarray}
A polynomial form of degree $n$ for $H (x)$ is achieved when we impose requirement that the series terminates \cite{ERFM,AVV,JM}. For this we must have
\begin{eqnarray}
&&\tilde{\beta}=\epsilon\,n \quad (n=1,2,3,.....),\nonumber\\
&&c_{n+1}=0.
\label{dd14}
\end{eqnarray}
Using the energy quantization condition $\tilde{\beta}=\epsilon\,n$, we get the following relation : 
\begin{equation}
\frac{\tilde{b}}{4}+B^2-2\,D\,(1 + 2\,A) = 4\,n\,D.
\label{dd15}
\end{equation}
Substituting $A, B, D$ using Eq. (\ref{aa24}) into the above equation (\ref{dd15}), we get the following energy eigenvalues:
\begin{equation}
E^{2}_{n,l}=k^2+m^2+\frac{16}{75}\,[\alpha\,(2\,n+2+\frac{l}{\eta})+\sqrt{\alpha^2\,(2\,n+2+\frac{l}{\eta})^2-\frac{15\,m^2\,(\Omega^2+\omega^2_{c})}{4\,\alpha^2}}]^2.
\label{dd16}
\end{equation}
It is clear from above that the energy eigenvalues of a charged particle depend on the parameter $\alpha$, oscillator frequency $\Omega$, external field $B_0$, and the topological defects $\eta$.

The eigen functions of the system in the presence of topological defects are
\begin{eqnarray}
\Phi_{n,l}(x)&=&x^{\frac{1}{2}(1+\frac{|l|}{\eta})}\,e^{\frac{x}{4\sqrt{-d}}\,(c+d\,x)}\,H (x),\nonumber\\
&=&x^{\frac{1}{2}(1+\frac{|l|}{\eta})}\,e^{\frac{x\,\alpha}{8}(1-\frac{x\,\alpha^2}{6})\,\sqrt{3\,(E^{2}_{n,l}-k^2-m^2)}}\,H (x),
\label{dd17}
\end{eqnarray}
where $E_{n,l}$ is given in Eq. (\ref{dd16}). By employing the additional recurrence condition $c_{n+1}=0$, one can evaluate the energy eigenvalues and corresponding eigen functions one by one as done in Refs. \cite{AVV,JM}.

For $\alpha \rightarrow 0$, the study space-time (\ref{2}) reduces to cosmic string space-time. The Klein-Gordon oscillator equation (\ref{dd8}) in this case becomes 
\begin{equation}
\psi'' (r)+\frac{1}{r}\,\psi' (r)+[\tilde{a}-m^2\,(\Omega^2+\omega^2_{c})\,r^2-\frac{{\bar l}^2}{r^2}]\,\psi (r)=0.
\label{dd18}
\end{equation}
Transforming a new variable $m\,\sqrt{\Omega^2+\omega^2_{c}}\,r^2=x$ into the Eq. (\ref{dd18}), we obtain the folloing equation \cite{AF}
\begin{equation}
\psi'' (x)+\frac{1}{x}\,\psi' (x)+\frac{1}{x^2}\,(-\xi_{1}\,x^2+\xi_{2}\,x-\xi_3)\,\psi (x)=0,
\label{dd19}
\end{equation}
where
\begin{equation}
\xi_1=\frac{1}{4}\quad,\quad \xi_2=\frac{\tilde{a}}{4\,m\,\sqrt{\Omega^2
+\omega^2_{c}}}\quad,\quad \xi_3=\frac{{\bar l}^2}{4}.
\label{dd20}
\end{equation}
The energy eigenvalues is given by
\begin{equation}
E_{n,l}=\pm\,\sqrt{k^2+m^2+2\,m\,\Omega+2\,m\,\omega_c\,\frac{l}{\eta}+2\,m\,\sqrt{\Omega^2+\omega^2_{c}}\,(2\,n+1+\frac{|l|}{\eta})},
\label{dd21}
\end{equation}
where $n=0,1,2,3,.........$. Equation (\ref{dd21}) is the energy eigenvalues of a Klein-Gordon oscillatory particle of frequency $\Omega$ in the presence of an external uniform magnetic field ($B_0$) in a cosmic string space-time.

A special case corresponds to $B_0 \rightarrow 0$, absence of an external uniform magnetic field into this relativistic system. The energy eigenvalues (\ref{dd21}) reduces to
\begin{equation}
E_{n,l}=\pm\,\sqrt{k^2+m^2+4\,m\,\Omega\,(n+1+\frac{|l|}{2\,\eta})}
\label{dd22}
\end{equation}
Equation (\ref{dd22}) is the energy eigenvalues of a Klein-Gordon oscillatory particle of frequency $\Omega$ in the background of cosmic string space-time without external fields. This eigenvalues is similar to the result obtained in Ref. \cite{Santos} without non-inertial effects (substituting $\omega \rightarrow 0$ into the Eq. (35) in Ref. \cite{Santos}).

Thus we can see that the energy eigenvalues (\ref{dd21}) of a Klein-Gordon oscillatory particle in the pesence of external fields in comparison to the result obtained in Ref. \cite{Santos} without non-inertial effects get modifies.

\vspace{0.1cm}
{\bf Case 2} : We choose the function $f(r) =a+b\,r^2$.
\vspace{0.1cm}

Therefore, Eq. (\ref{dd7}) becomes 
\begin{equation}
\psi'' (r)+\frac{1}{r}\,\psi' (r)+(\tilde{a}+\tilde{b}\,r^2-\frac{{\bar l}^2}{r^2})\,\psi (r)=0.
\label{cc1}
\end{equation}
Transforming a new variable $r^2=x$ into the Eq. (\ref{cc1}), we have the following equation \cite{AF}
\begin{equation}
\psi'' (x)+\frac{1}{x}\,\psi' (x)+\frac{1}{x^2}\,(-\xi_{1}\,x^2+\xi_{2}\,x-\xi_3)\,\psi (x)=0,
\label{cc2}
\end{equation}
where
\begin{equation}
\xi_1=-\frac{\tilde{b}}{4}\quad,\quad \xi_2=\frac{\tilde{a}}{4}\quad,\quad \xi_3=\frac{{\bar l}^2}{4}.
\label{cc3}
\end{equation}
The energy eigenvalues is given by
\begin{eqnarray}
&&E^{2}_{n,l}=k^2+m^2+2\,m\,(\Omega+\omega_c\,{\bar l})+\alpha^2\,(2\,n+1+|{\bar l}|)^2\nonumber\\
&&+(2\,n+1+|{\bar l}|) \{\alpha^4\,(2\,n+1+|{\bar l}|)^2+(2\,m\,\Omega+\alpha^2)^2\nonumber\\
&&+4\,m\,\omega_c\,(m\,\omega_c+\alpha^2\,{\bar l})-\alpha^4\}^{\frac{1}{2}},\quad\quad
\label{cc4}
\end{eqnarray}
where ${\bar l}=\frac{l}{\alpha}$, an effective angular momentum eigenvalues. The energy eigenvalues depend on the parameter $\alpha$, the external field $B_0$, the topological defects $\eta$, and the oscillator frequency $\Omega$ of the particle. 

For $\alpha \rightarrow 0$ only, the study curved space-time reduces to cosmic string space-time. The energy eigenvalues (\ref{cc4}) reduces to the result (\ref{dd21}). For zero oscillator frequency only, $\Omega \rightarrow 0$, the energy eigenvalues (\ref{cc4}) reduces to the result Eq. (\ref{bb5}). 

For $\alpha \rightarrow 0$ and the absence of external field $B_0 \rightarrow 0$, the energy eigenvalues (\ref{cc4}) reduces to the result obtained in Ref. \cite{Santos} in a cosmic string space-time without non-inertial effects.

The wave functions are given by 
\begin{equation}
\psi_{n,l}=|N| x^{\frac{|l|}{2\,\eta}}\,e^{-\frac{x}{2}\,\sqrt{m^2\,\Omega^2+\frac{\alpha^2}{2}\,(E^2_{n,l}-k^2-m^2)}}\,L^{(\frac{|l|}{\eta})}_{n} (\sqrt{m^2\,\Omega^2+\frac{\alpha^2}{2}\,(E^2_{n,l}-k^2-m^2)}\,x).
\label{cc5}
\end{equation}

From the above analysis by {\bf Case 1-Case 2}, we can see that the energy spectrum Eq. (\ref{dd21}) and (\ref{cc4}) of a charged particle in the presence of an external uniform magnetic field in a curved space-time modifies the result obtained in Ref. \cite{Santos} in a cosmic string space-time without non-inertial effects (substituting $\omega \rightarrow 0$ into the Eq. (35) in Ref. \cite{Santos}).

\section{Conclusions}

Cosmic strings are hypothetical massive objects that may have contributed to the anisotropy of the cosmic microwave background radiation and, consequently, to the large scale structure of the Universe \cite{AL}. There existence is also supported in superstring theories with either compactified or extended extra dimensions. Both static and rotating cosmic strings can be equally responsible for some effects such as particle self-force \cite{ERBM2,CRM2} and gravitational lensing \cite{MVS}, as well as for the production of high energetic particles \cite{VBB3,VAL,JA3}. Notably, out of the conical singularity, cosmic string present a flat space-time geometry ($R_{\mu\nu\rho\sigma}=0$) with some remarkable global properties. These properties include theoretically predicted effects such as gravitomagnetism and (non-quantum) gravitational Aharonov-Bohm effect \cite{AB2,VBB4}.

The relativistic energy eigenvalues of scalar particles in G\"{o}del and G\"{o}del-type metrics with or without cosmic string has been addressed by several authors. They demonstrated that the energy eigenvalues of the system modifies and depend on the parameters characterizing the space-time. In Ref. \cite{ERFM}, fermionic and bosonic field in the presence of external fields with scalar potential in a cosmic string space-time, were investigated and obtained the energy eigenvalues. They have shown that the energy eigenvalues in the presence of exteral uniform magnetic field get modifies in comparison to the Landau levels. In Ref. \cite{Santos}, relativistic quantum dynamics of scalar particle under the influence of non-inertial effects in the cosmic string space-time without any scalar potential, were investigated. They solved the KG-oscillator in the background space-time generated by a cosmic string without any potential, and obtained the relativistic energy eigenvalues. Based on the above two works Refs. \cite{ERFM,Santos}, in this article, we study relativistic quantum dynamics of spin-$0$ particles in the presence of an external uniform magnetic field in a curved space-time which is the generalization of cosmic string metric.

In {\it sub-section 2.1}, we have studied the relativistic quantum motion spin-$0$ massive charged particle in the presence of an external uniform magnetic field in a curved space-time. For the function $f(r)=a +b\, r^2+c\, r^4+d\, r^6$ into the radial wave-equation, we have obtained the energy eigenvalue Eq. (\ref{aa34}) which get modifies in comparsion to the result obtained in Ref. \cite{ERFM} in a cosmic string space-time. The eigenvalues depend on the parameter $\alpha$, external magnetic field $B_0$, and the topological defect $\eta$. Similarly, for the chosen function $f(r)=a+b\, r^2$ where $\alpha < < 1$, we have obtained the energy eigenvalues Eq. (\ref{bb5}). We have seen that for $\alpha \rightarrow 0$, the energy eigenvalues (\ref{bb5}) reduces to the result obtained in Ref. \cite{ERFM} in a cosmic string space-time. Thus the energy spectrum of spin-0 massive charged particle in the presence of external fields in a curved space-time are the modified result in comparison to the case  obtained in Ref. \cite{ERFM} in a cosmic string space-time (see Eq. (23) in Ref. \cite{ERFM}).

In {\it sub-section 2.2}, we have studied the Klein-Gordon oscillator in the presence of external uniform magnetic field in the background of curved space-time with topological defect. For the chosen function $f(r)=a +b\, r^2+c\, r^4+d\, r^6$ into the radial wave-equation, we have evaluated the energy eigenvalues Eq. (\ref{dd16}) and corresponding eigenfunctions (\ref{dd17}). The eigenvalues depend on the parameter $\alpha$, the oscillator frequency $\omega$, the external field $B_0$, and the topological defect $\eta$. We also inevstigated for $\alpha \rightarrow 0$ there and obtained the energy eigenvalues (\ref{dd21}) for a massive charged particle in the presence of external field and compare with the result obtained in Ref. \cite{Santos} in a cosmic string space-time without non-inertial effects. Similarly for the function $f(r)=a +b\, r^2$, we have obtained the energy spectrum Eq. (\ref{cc4}) and the corresponding wave-functions Eq. (\ref{cc5}). For $\alpha \rightarrow 0$ and absence of the external fields $B_0 \rightarrow 0$, the energy eigenvalues Eq. (\ref{cc4}) reduces to the result obtained in Ref. \cite{Santos} in a cosmic string space-time without non-inertial effects. Thus the energy spectrum for spin-$0$ massive charged particles in the presence of external fields in a curved space-time modifies the result obtained in Ref. \cite{Santos} without non-inertial effects (substituting $\omega \rightarrow 0$ into the Eq. (35) in Ref. \cite{Santos}).

Before we finish this paper, we would like to mention two potential applications of our results along with the result obtained in Ref. \cite{ERFM}. They are in Condensed Matter Physics and Astrophysics. As to Condensed Matter Physics ststem \cite{CS,AMMC,CS2,CS3}, it is well known that the linear defect in elastic solid named disclination can be dealt with in the same geometric approach as to the cosmic strings \cite{HM,FM,MOK2,CF2}. With respect to the astrophysical application, this analysis may also be useful to understand the quantum motion of charged particles in the Cosmos under the influence of galactic magnetic field considering the presence of a cosmic string \cite{MB}. Besides topological defects like cosmic strings \cite{AV}, domain walls \cite{AV2}, and a global monopole \cite{MB2} provides a tiny relation between the effects in cosmology and gravitation. In fact, the analysis developed here constitutes a relativistic extension of the result in Ref. \cite{ERFM} to study the quantum mechanical motion of charged particles under the influence of external uniform magnetic field in the curved space-time with a cosmic string.

Summarizing, we can say that, by the results obtained in this paper, the presence of topological defect (cosmic string) $\eta$, the parameter $\alpha$ (which causes Riemannian flat cosmic string geometry into a curved one) as well as the external uniform magnetic field $B_0$ produces a significant modification on the energy spectra in comparison to the results obtained in Ref. \cite{ERFM}, and in Ref. \cite{Santos}. So in this paper, we have shown some results for quantum systems where the general relativistic effects are taken into account, which in addition to the previous known results in Refs. \cite{ERFM,Santos} present many interesting effects. This is a fundamental subject in physics, and the connections between these theories are not well understood.



\end{document}